\documentclass[review]{elsarticle}

\usepackage{graphicx,color}
\begin{document}

\title{Enhancing retinal images by nonlinear registration}

\author[weizmann,lesia]{G.÷Molodij\corref{cor1}}
\ead{guillaume.molodij@obspm.fr}
\author[technion]{E.N.÷Ribak}
\ead{eribak@physics.technion.ac.il}
\author[lesia]{M.÷Glanc}
\ead{marie.glanc@obspm.fr}
\author[Vision,lesia]{G.÷Chenegros}
\ead{guillaume.chenegros@obspm.fr}
\cortext[cor1]{Corresponding author}
\address[weizmann]{Faculty of Physics, Weizmann Institute of Science, P.O. Box 26, Rehovot 76100, Israel}
\address[lesia]{LESIA-Observatoire de Paris-Meudon, CNRS, Universit\'e Pierre et Marie Curie-Paris 06, Universit\'e Paris Diderot - Paris 07, 5 place J.Janssen, 92190 Meudon}
\address[technion]{Department of Physics, Technion -- Israel Institute of Technology, Haifa 32000, Israel}
\address[Vision]{Institut de la Vision UMR-S 968, Inserm, UPMC, CNRS 7210, CHNO des Quinze-Vingts, 75012 PARIS}

\date{Received; revised ; accepted}

\begin{abstract}
Being able to image the human retina in high resolution opens a new era in many important fields, such as pharmacological research for retinal diseases, researches in human cognition, nervous system, metabolism and blood stream, to name a few. In this paper, we propose to share the knowledge acquired in the fields of optics and imaging in solar astrophysics in order to improve the retinal imaging at very high spatial resolution in the perspective to perform a medical diagnosis. The main purpose would be to assist health care practitioners by enhancing retinal images and detect abnormal features. We apply a nonlinear registration method using local correlation tracking to increase the field of view and follow structure evolutions using  correlation techniques borrowed from solar astronomy technique expertise.  Another purpose is to define the tracer of movements after analyzing local correlations to follow the proper motions of an image from one moment to another, such as changes in optical flows that would be of high interest in a medical diagnosis.

\end{abstract}

\maketitle

\section{Introduction}

High-resolution imaging of the retina has significant importance for science: physics and optics, biology, and medicine \cite{Li07,Chui08,Uchida13}. Early detection of retinal pathologies can be performed by noninvasive imaging of the retinal tissue to a cellular level. However, the eye is difficult to observe because it presents static and chromatic aberrations. In addition, the live observation of the retina is perturbed by dynamic aberrations of the eye which can be compensated, in real time, by an adaptive optics system \cite{Miller96,Liang97,Drexler01,Glanc02,Glanc04}. This correction is however partial and resolution of the images obtained is not sufficient \cite{Roggemann91}. To circumvent this problem, images are mosaiced from scanning by confocal lasers or optical coherence tomography.

The observation of abnormal cells patterns of the ill retina is accompanied by unfavorable conditions that make the analysis more difficult. Patients with severe deformities of the cornea can no longer fit in the analysis sessions because of the difficulty to focus on their retinae. Images are thus of low contrast, and difficult to interpret. Further work, and major challenge for medical applications, should be to assess the ability of processing methods of retinal images to function in very poor conditions with a reliability rate of the highest possible standards. In a previous study, we explored a new aspect of the analysis of image quality degradation based on structural information and we demonstrated enhanced processing of sequences of fundus images obtained using a commercial adaptive optics flood illumination system \cite{Molodij14}. In the case of the retinal images considered, we process sequences of images with random translations produced by ocular saccades. The illumination component is determined from the sum of the images in the sequence as it retains information on large structures in the retina such as vessels and have to be analyzed in bad conditions for wide fields of view at the expense of the resolution.

The selection and the re-centering of large field of view solar images is a technique which has been applied to astronomical images for several years and is referred to local correlation tracking technique (LCT) originally dedicated to the elimination of distortions coming from terrestrial atmospheric turbulence on images of the solar granulation \cite{November86,November88,Molodij10}. This method is able to correct aberrations due to eye movements during measurements on a basis of a nonlinear registration (NLR) approach, and increase the entire field of view that can be explored at high resolution taking into account  the image shifts as well as the the image distortion and rotation \cite{Glasbey98, Jain98, Redert99, Uchida05}. The tracer of movements after analyzing local correlations can also follow the proper motions of an image from one moment to another, that would be of high interest in a medical diagnosis. The template is fixed or updated in a frame-by-frame manner or use various criteria for evaluating the similarity to follow the change of the target appearance. 

Nonlinear image registration is equivalent to optical flow because optical flow is usually applied to a pair of consecutive video frames as the image registration.  The optical flow is a technique to estimate the motion at all the pixels at each frame to visualize the motion field on the entire image \cite{Horn81,Beauchemin95}. The technique is useful to analyze not only the motion but also the proper deformation in the tracked field of view. The result of object tracking, i.e. the temporal trajectory of the target, is often used for plotting a velocity map and a moving direction histogram. It is also useful to analyze some other characteristics, such as a deviation from a reference. The optical flow technique is useful to analyze not only the motion but also the deformation of flexible objects. Another function would be to evaluate the similarity or dissimilarity between two temporal patterns and of interest for health care practitioners. Many object tracking methods have been proposed \cite{Yilmaz06}. Nevertheless, the visual object tracking in bioimages is still an open problem and thus is very challenging. 

Adaptive optics corrected image sequences show variations in the quality that deteriorates before and during blinks of the eye, but variations are also observed although not associated with blinks. Nevertheless, the use of a multiframe mode with a registration method to increase the signal-over-noise ratio must incorporate an image selecting step to exclude poor quality images from further processing as well as to examine the extent of the final improved image quality. For this purpose, we need a reliable measure of image quality, and many methods have been proposed for determining image quality. In a previous work, attempts have been made to measure the quality of extended fundus images \cite{Molodij14}. These images have less high-resolution details, and typically involve measuring the size of the edges of large anatomical features such as blood vessels \cite{Fleming06}.
 
In this paper, we present an image processing algorithm for resolution enhancement of extended retinal images. The novelty in the suggested method is the ability to significantly improve the resolution of an ensemble of poor quality and large field of view retinal images using a nonlinear registration method based on the LCT technique. We recall briefly in Section 2 the technique to determine the proper motion that maximizes the spatially localized cross correlation between two images of a scene separated by a sampling time shorter than the lifetime of tracers in the scene. In Section 3, we compare and discuss the different image quality assessments to determine the improvement in the field of view. Section 4 is dedicated to the result obtained with real data. We discuss the application of this study in the fields of optics and imaging in astrophysics in order to improve the visualization of the retina at very high spatial resolution, and to implement methods for high-resolution retinal imaging purposes.

\section{Nonlinear registration}

Image registration is the technique to fit an image to another image. The simplest linear transformation function are horizontal and vertical shifts. Rotation and scaling are also typical linear geometric transformations for image registration. A more general geometric transformation is the affine transformation that includes translation, rotation, scaling, shear, and their arbitrary combinations. Nonlinear image registration, or deformable template, realizes more flexible image registration that can map a straight line as a curve.

Registration can be broadly classified into intensity or feature based. The intensity based methods have the inconvenience of poor performance under varying illumination while feature methods are based on accurate and repeatable extraction of the features. Usually, intensity based methods are applied on retinal images which may not contain sharp information \cite{Xue07, Ramaswamy13}. Nevertheless, we showed in a previous work that retinal images contain enough details to extract structural information for direct template matching, without the need to detect prominent features \cite{Molodij14}. 

\begin{figure}
\centering
\includegraphics[width=11.cm]{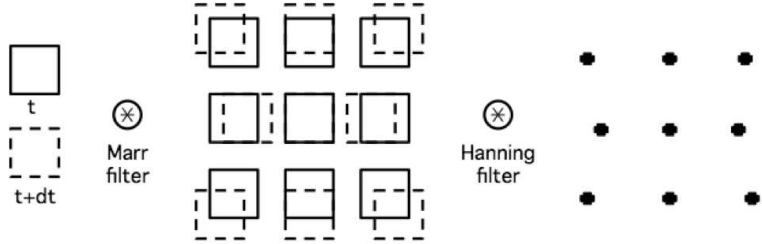}
\caption{Schematic representation of the numerical method to determine the local cross correlation and the displacement map. The original sub raster image obtained at time $t$ and $t+ dt$ is filtered using the convolution with a Marr-Hildreth filter. In this example, nine subfields allow the determination of the maximum of the cross correlation using a fast Fourier algorithm and using the Hanning filter. The position are collected into a table of control (tile) points. }
\label{schem.eps}
\end{figure}

Fig. \ref{schem.eps} shows the principal steps of the registration method that is to tile the images with square kernels. For each kernels, a two dimensional cross correlation is computed with respect to a reference image convolved with a Marr-Hildreth filter \cite{Marr80} in order to overcome the aperture edge discontinuities which are lost, especially when the kernel becomes smaller compared to the template spatial autocorrelation. This filter is analogous to the commonly used 'difference of gaussians', where the Laplacian operator is approximated by the difference of blurred versions of the object.

The position of the maximum is filtered using Hanning filter \cite{Molodij96}, and interpolated for sub-pixel positions. The positions are collected into a table of control (tile) points after removing the eye saccade displacements from the smooth drifts in respect with the sampling. A b-spline surface is fitted to these tile points \cite{Foley82}. Finally, the scene is resampled on the nonuniform grid. 
The distortion mapping measured in these data is a continuous function of space whose length scale of variation is generally larger than the width of the window function. 

\begin{figure}
\centering
\includegraphics[width=12.cm]{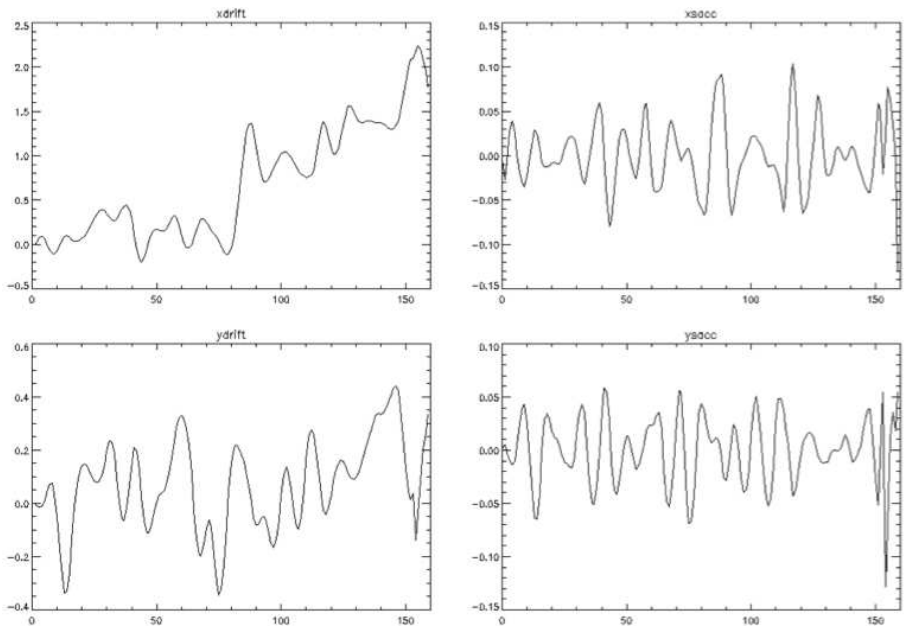}
\caption{The plots display the respective contributions of the smooth drifts and the eye saccades (indicated in pixels) along x and y directions of one of the control points for each of the 160 images of the sequence (indicated x on the axis). The fit removing eye-saccades are in agreement with the assumption that the residual motion is less than the typical scale of the features. }
\label{nlrdrift.eps}
\end{figure}

We determine the vector displacement that maximizes the spatially local cross-correlation between an image and a reference image as a continuous function of the image space. This technique was first applied to a long temporal sequence of the solar observations of granulation made during a condition of rapidly changing atmospheric distortion wherein features were remapped without loss of spatial resolution \cite{November86}. Fig. \ref{nlrdrift.eps} shows the resulting amplitude of displacement as a function of distance from the fixed point averaged over 160 distortion maps generated by comparing an image to an ongoing time average of images. The relatively long temporal sequence is a necessary condition to weight the contribution of the distortion in the image in order to remove large eye saccade displacements from the smooth drifts compared to the sampling. The local displacement is a nonlinear function of the spatially localized cross correlations. The temporal average of the spatially localized correlation is affected in a minimum way by the poor-displacement contributions. The effect of eye saccades is large enough so it is important that its contribution be minimized in the average of many measurements. The instantaneous amplitude of the geometric distortion due to saccades is larger than the amplitude of the flows we wish to measure. When the image quality is poor and the contrast low, the cross-correlation peak is broad and is a nearly constant function of displacement. The poor-blurring contribution to the time-averaged cross correlation does not bias its centroid strongly and so provides a natural method for weighting contributions taken during varying conditions. The quantification of proper-motion measurements have been validated from space-based observations with the SOUP experiment using the cross correlation technique \cite{Title89}. 

\section{Image quality assessments}
\subsection{Image quality metrics}
Many methods have been proposed for determining image quality \cite{Ramaswamy13}. We recall briefly the characteristics of the different methods investigated in the paper.
\subsubsection{Image variance}
A higher variance implies a sharper image. Nevertheless, a noisier image will also present higher variance.  The image intensity variance is
\begin{equation}
\sigma^2 = \frac{1}{n-1} \sum_{i,j} \left[ I(i,j) - \bar{I}\right]
\end{equation}
Where $n$ is the number of pixels in the image and $\bar{I}$ is the mean pixel brightness in each image. The sum is taken over the whole image, or a window in the region of interest.
\subsubsection{Image contrast}
The simplest and most widely used full-reference quality metric is the image contrast, computed by,
\begin{equation}
(\frac{\Delta I}{\bar{I}})_{r.m.s.}=\sqrt{\frac{\sum_{i}^{n}(I_{i}-\bar{I})^{2}}{n\bar{I^{2}}}}  \label{contraste}
\end{equation}
where $\bar{I}$ is the average pixel value for the given image.
\subsubsection{Entropy and redundancy}
Rather than searching for features in an image, the visual system codes a given image with regard to the statistical properties of the set of natural images \cite{Kersten87}. Because the space of possible pictures is so great, it makes good sense to utilize naturally occurring redundancy to recode image information into a less redundant that was originally defined by Shannon \cite{Shannon51}. Let an image be specified by $k$ pixels with $m$ bits of gray-level resolution per pixel. The n$^{th}$ order
conditional entropy for this class of pictures, $F_n$, is the expected value of the negative $\log$ (base 2) of the probability of gray level $i$ conditional on the values of $n$ neighbors (over some defined neighborhood structure):
\begin{equation}
F_n = - \sum_{i,j} p(i,b_j) \log_2 p(i |b_j)
\end{equation}
where $b_j$ is the $j^{th}$ block of the $n$ neighborhood pixels ($j$ = 1 to 2$^{mn}$, $i$ = 1 to 2$^m$). As $n$ approaches $k$, $F_n$ approaches the minimum average number of bits per pixel required to code this class, for arbitrarily small error. If the probability of pixel gray levels is constant and independent of all others, the entropy is a maximum value of $m$ bits per pixel. The redundancy is then, $1 - \frac{F_n}{m}$. The probabilities are approximated by the histogram of the image. An interesting application of the entropy method to find the optimal window for image analysis can be determined by means of  the information theory \cite{Giammanco00}.
\subsubsection{Kurtosis}
The kurtosis is a statistical measure of the peakedness or flatness of a distribution and can be used as a measure of sharpness. Kurtosis is defined as 
\begin{equation}
\gamma_2= \frac{\mu_4}{\mu_2^2} - 3
\label{Kurto}
\end{equation}
where $\mu_4$ and $\mu_2$ are the fourth and second central moments. The central moment of order $n$ is 
\begin{equation}
\mu_n = \int_{-\infty}^{+\infty} (x-\bar{x})^n I(x) \textrm{d}x
\end{equation}
For a Gaussian distribution $\gamma_2$ = 3, therefore a non-Gaussian distribution has a non-zero kurtosis. In practice, we do not subtract the Gaussian distribution in Eq. \ref{Kurto} and we plot $\gamma_2 + 3$ in Fig. \ref{qicomp.eps}
\subsubsection{Sharpness}
The image sharpening approach was proposed as an adaptive correction of astronomical images \cite{Muller74}.  A designer metrics which could emphasize the contrast of both bright points and dark areas was given by \cite{Fienup03}.
\begin{equation}
S_r = \sum_{i,j} w(i,j) \left[ I(i,j) - \gamma \bar{I}\right]^{\beta},
\end{equation}
where $w(i,j)$ is a weighting function that can be used to mask pixel artifacts, $n$ is the number of pixels in the image and $\bar{I}$ is the mean pixel brightness in each image. Nevertheless,  the parameters ($\beta, \gamma$) of the metric depend strongly on the content of the scene, with large powers of $\beta$ best for images containing prominent points and small powers of $\beta$ best for images containing dark regions with no prominent points. \\
The simplest metric (with $\beta = 2$ and $\gamma = 0$) is defined by
\begin{equation}
S_r = \sum_{i,j} w(i,j) \left[ I(i,j)\right]^2
\end{equation}
gives also a sharpening assessment adapted to the retinal images  when the mean intensity value is constant \cite{Zommer06}.
\subsubsection{Structural information assessment}
A modified Weber-Fechner criterion that takes into account the logarithmic sensitivity of eyes in terms of the light and the structural information in the image was proposed \cite{Molodij14}. Most images are highly structured in the sense that each pixel is dependent on its neighboring pixels. This dependence provides an information on the structure of objects in a scene. The structural information assessment is given by
\begin{equation}
A_{SI}  = 20 \, Log_{10} \frac{Max[I_r^2(i,j)]}{\delta_g}, \label{WF}
\end{equation}
\noindent with
\begin{equation}
\delta_g = \frac{1}{n} \sum_{i,j} \left[\nabla I_r(i,j) - \nabla I_c(i,j)\right]^2 , \label{superweber}
\end{equation}
\noindent where $n$ is the number of pixels in the image. The gradient is taken at half scale of the image spatial correlation.

\subsection{Comparison}
\begin{figure}
\centering
\includegraphics[width=12.cm]{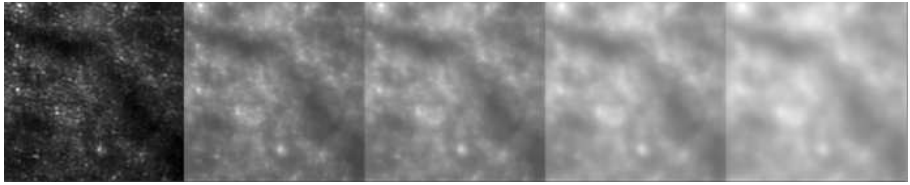}
\caption{Simulation of an increasing stochastic blurring determined with OTFs derived from the Zernike expansion. The retinal image left on the figure is defined as the perfect reference image associated with a Strehl value of 1.0. By comparison, the Strehl ratio of the corresponding images left to right are: 0.23, 0.15, 0.065 and 0.03.}
\label{simu.eps}
\end{figure}

In order to evaluate the NLR improvement by comparison to linear registration method, we first  investigate the reliability of the image quality assessments using simulated optical transfer functions (OTFs). We compare metrics such as the image contrast, the entropy, the kurtosis, the sharpness, and the structural information assessment. We use simulations of stochastic blurring taking into account our knowledge on the effects of the earth's atmosphere on astrophysical images.  In the simulation, we estimate accurately the quality of the image using OTFs convolved with the retinal images. A Zernike expansion model following the Wang and Markey approach \cite{Wang-78} is used to derive the OTFs \cite{Molodij-98,Molodij-12}.  Fig. \ref{simu.eps} shows the blurred image obtained with the corresponding calculated OTFs. 

\begin{figure}
\includegraphics[width=12.cm]{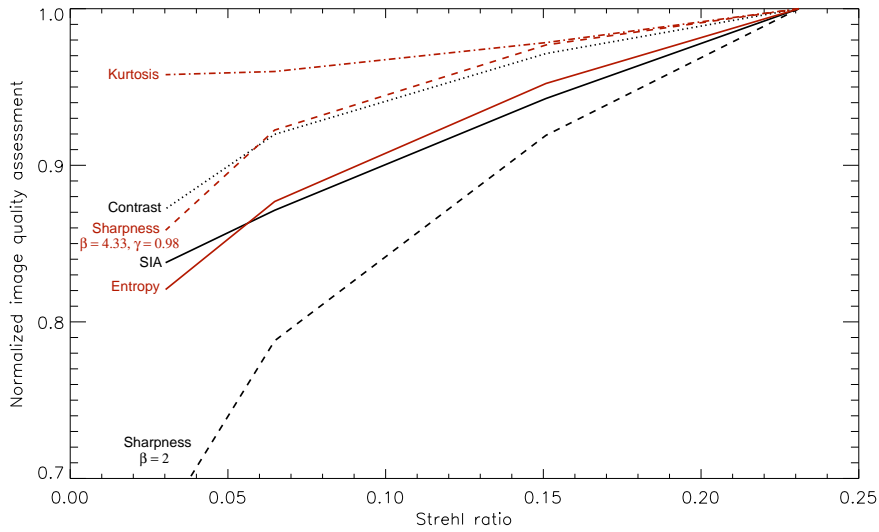}
\caption{ Comparison between the normalized image quality assessments  versus the Strehl ratio. In red curves are plotted the most interesting absolute (without calibration) assessments such as the entropy, the kurtosis and the sharpness with parameters $\beta$=3.66, $\gamma$=0.98. Sharpness and contrast methods show very close behavior. Despite its valuable linearly dependence on the Strehl ratio, the structural information assessment (SIA) needs a reference template. Kurtosis shows a good linear behavior but a lack of sensitivity. Dashed black line corresponding to sharpness with parameters $\beta$=-2.0, $\gamma$=0 appears to be too sensitive to the mean image variations. We find that entropy assessment seems to match better with our subjective estimation of the quality, showing interesting close linear property in respect with the image corrugation.}
\label{qicomp.eps}
\end{figure}

Fig.\ref{qicomp.eps} displays the image quality assessments for an increasing effect of the corrugation applied on the same template. The reference image (left on Fig. \ref{simu.eps}) is considered as the perfect reference both for the normalizations and the derivation of the structural information assessment (SIA).  In Fig. \ref{qicomp.eps}, the SIA, the entropy and the kurtosis methods show a close linear behavior with respect to the Strehl ratio and lead to a more precise evaluation of the image quality assessment. The kurtosis method has the great advantage to be self-calibrated (a Gaussian distribution $\gamma_2$ = 3) as well as the entropy on the basis of the grey levels. By comparison, the SIA criterion needs a reference image and is more adapted to the evaluation of the difference between images. The sharpness parameters ($\beta$ and $\gamma$) are very sensitive-keys and depend on the content of the scene. The $\gamma$ parameter purpose is essentially  to limit the effect of the mean intensity on the metric. Sharpness with parameters $\beta$=3.66, $\gamma$=0.98 and contrast methods show very close behavior. Nevertheless, we notice the important effect of the detector dust (removed by the registration mean summation) and the template appearance modifications du to eye-saccades on the sequence for kurtosis and sharpness assessments.  We find that entropy assessment seems to match better with our subjective estimation of the quality. Entropy method shows also interesting close-linear property in respect with the image corrugation.

\section{Results}

In order to test the method, we developed software in the Interactive Data Language (IDL) environment. We performed the processing method on several sets of data obtained at the XV-XX Hospital and at the Ichilov hospital, Tel Aviv. LESIA at the Paris Observatory has invested in this technology and has constructed an optical bench with a retinal imaging adaptive optics system at the XV-XX Hospital, Paris \cite{Glanc02,Glanc04}. 
\begin{figure}
\centering
\includegraphics[width=11.cm]{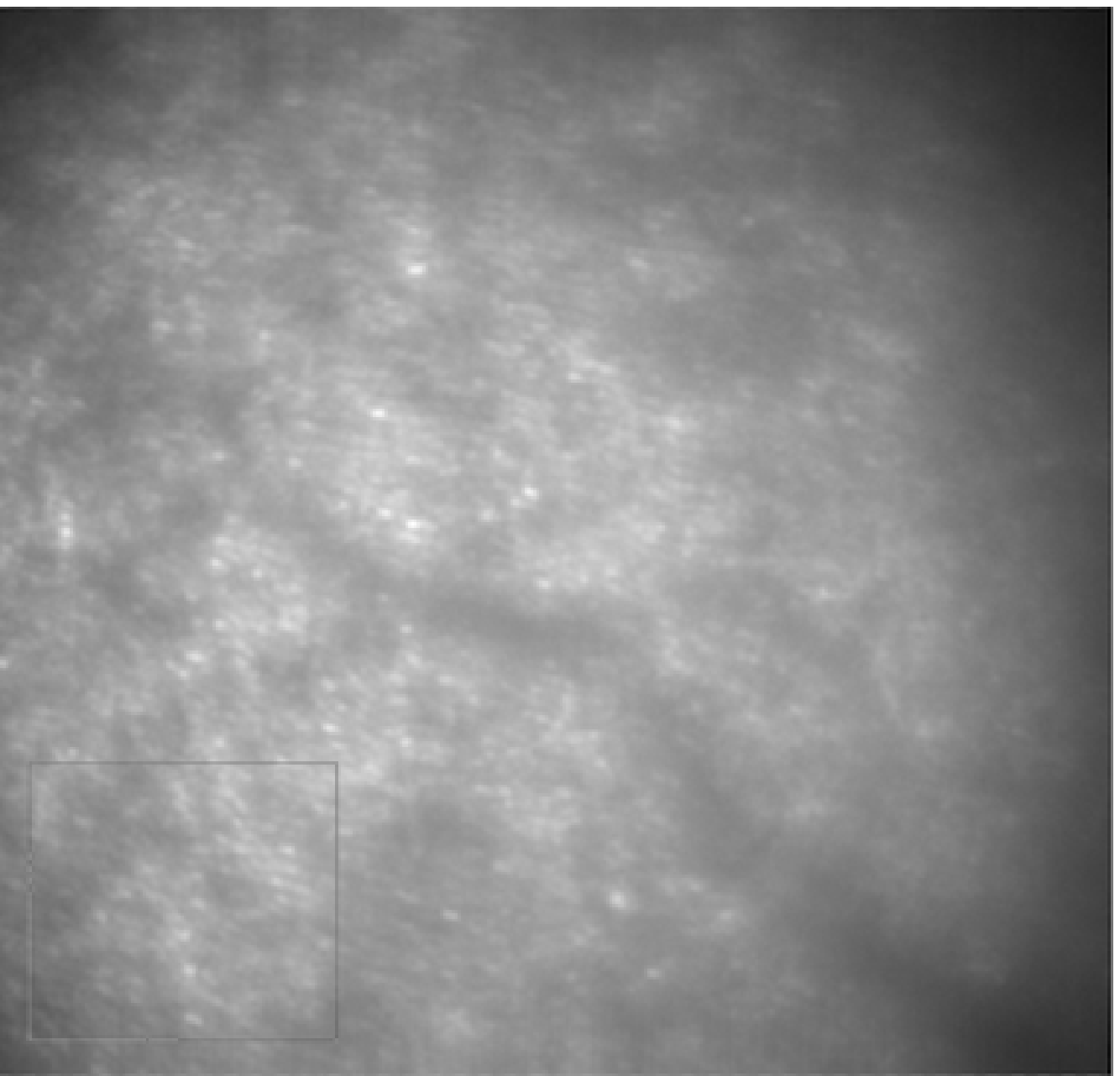}
\includegraphics[width=5.5cm]{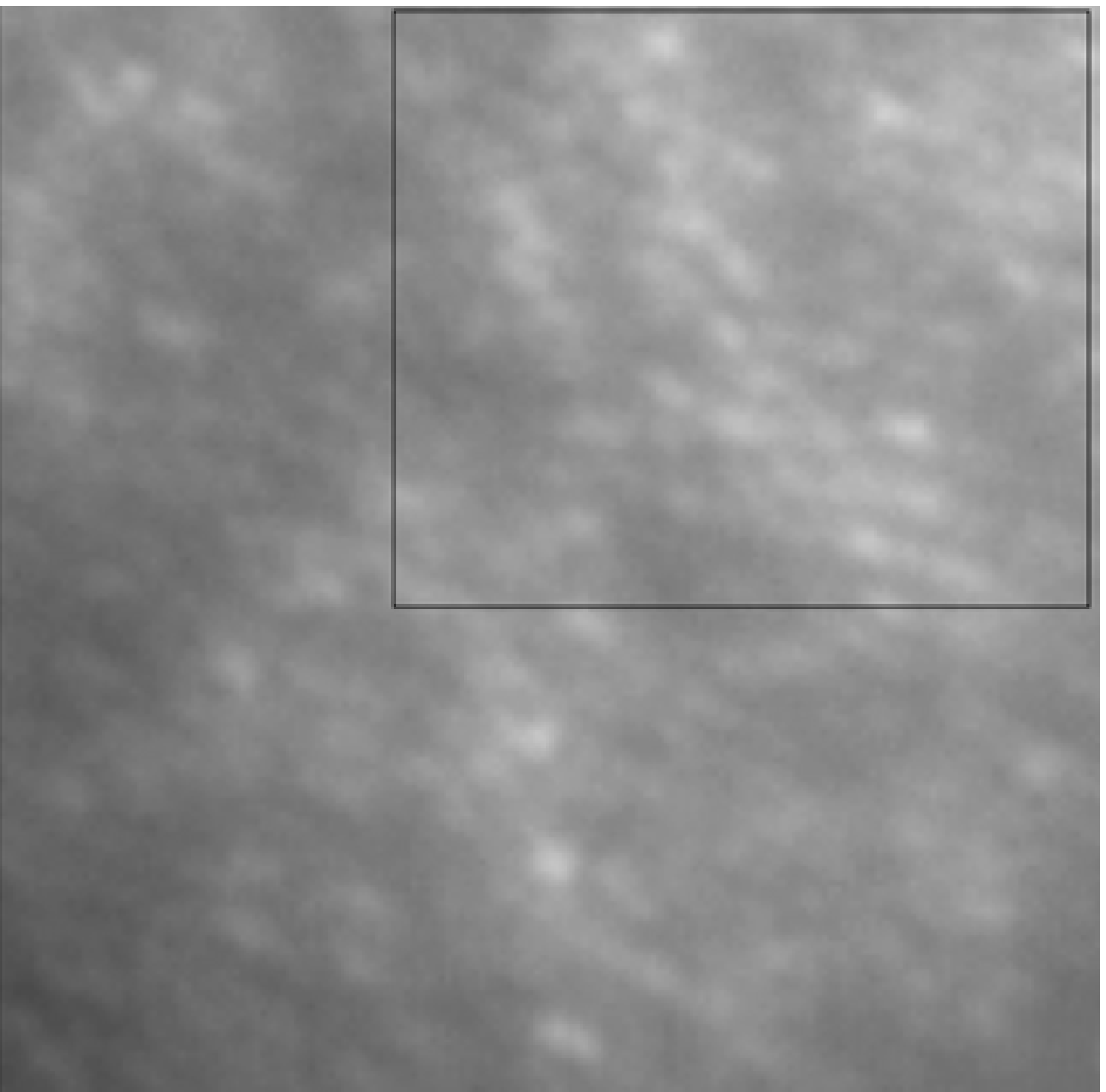}
\includegraphics[width=5.5cm]{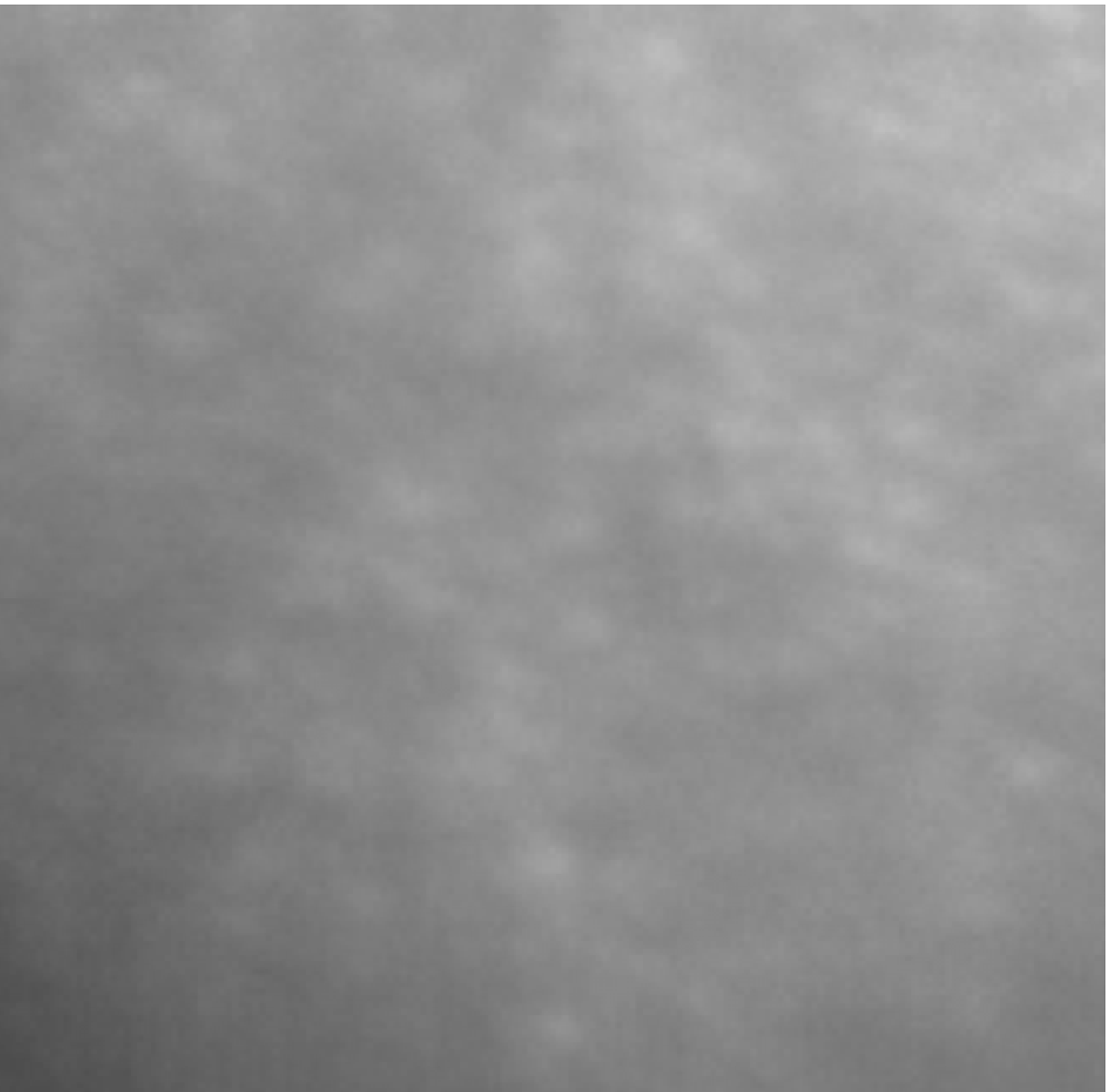}
\caption{Top: final registered retinal image using the proposed LCT method. Fifteen images are summed to derive a registration under the pixel precision.  Bottom: comparison between the same sub-field NLR (left) and linear registration (right) on a sequence of 50 images obtained at the XV-XX Hospital with the adaptive optics bench OEIL, 3ˆ January, 2009. The sub-raster is indicated (top).The image quality ratio between the nonlinear and linear methods in the sub-field is evaluated 1.07 by entropy, 1.10 by kurtosis while the sharpness gives a value of 0.95, i.e., a better image quality for the linear registration method.}
\label{nlr.eps}
\end{figure}
Fig. \ref{nlr.eps} displays the registration on fifteen high resolution images using the nonlinear method to correct the residual distortions after adaptive optics compensation. We evaluate the reliability of  the processing in a sub-field by comparison to the linear registration method developed in a previous work \cite{Molodij14}. The purpose is to maintain the highest resolution on a larger field of view. The registration is increasing the reliability to detect true features. The processing is applied on several sequences obtained with the optical bench "OEIL" at the XV-XX Hospital (overall 130 images observed 30 of January, 2009). The detector is a Q-Imaging CCD 6.45 micron pixel size at 550 nm wavelength. During the long sequence, only fifty images show a similar template that can be registered on a sufficient large field of view. 

As a prerequisite before the determination of the local two-dimensional cross-correlation, and in order to save the computational resources, we attempt to correct the large shifts coming from the rapid eye motion using the correlation method based on the complex conjugate maximum and the fast Fourier transform algorithms on the entire field of view. Then, we select the images according to similarity criterion ($\simeq$ 33 \% of the AO high resolution images), i.e., using the structural information assessment on the region of interest to select images  that have been correctly shifted under the assumption that the residual motion is less than the typical scale of the features \cite{Molodij14}. We find that this is quite robust when the template is a large sub-window of a high-quality reference image. 

\begin{figure}
\centering
\includegraphics[width=11.cm]{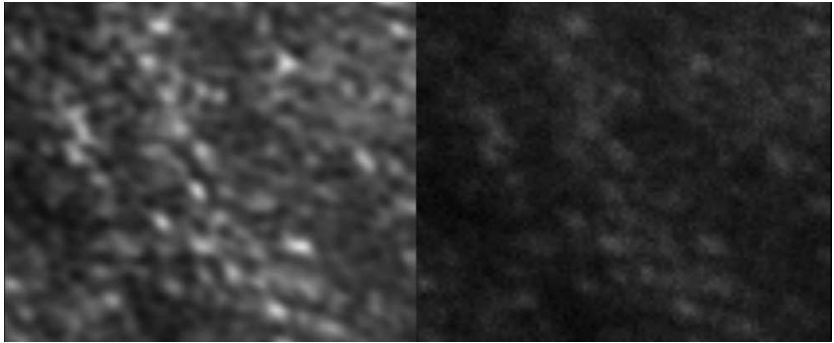}
\caption{Example of super-resolution (by a factor of four) on the sub-raster indicated bottom left on Fig. \ref{nlr.eps}. By comparison, the best raw image of the sequence is displayed right. The image quality ratio between the registered and the raw images is 1.67 by the entropy assessement. The same intensity scale is used for both images. }
\label{superesol.eps}
\end{figure}

We apply the entropy assessment on different part of the field to compare the image quality between the two methods. We define an image quality ratio based on entropy assessment between the linear and non linear method on the registration made up the same selected images. This ratio is close to 1 at the center part of the field and is around 1.07 at the corners of the field of view to indicate first the reliability of  the NLR method and a better efficiency far from the center part of the image where is applied the rigid rotation correction of the linear method. We conclude in the same way using the kurtosis assessment that indicates values from 1 to 1.1 from center to corners. Nevertheless, sharpness assessment gives a value of 0.95 on the entire field of view (i.e., in favor of the linear method). We find that entropy assessment seems to match better with our subjective estimation of the quality, showing interesting close linear property in respect with the image corrugation and, being a robust criterion also with significant changes of the template. Fig. \ref{superesol.eps} shows an example of super-resolution obtained after the registration on the selected images from the sequence. The resolution displayed is four times better by comparison to the best raw image.\\

Optical flow-methods can be broadly classified into correlation, feature or gradient based methods. The correlation based techniques compare parts of the first image with parts of the second in terms of the similarity in brightness patterns in order to determine the motion vectors. The feature-based methods compute and analyze optical flow at small number of well-defined images features while the gradient based methods use spatiotemporal partial derivatives to estimate flow at each point. Horn and Schunck  \cite{Horn81} introduced a regularity condition to minimize both the optical flow constraint (aperture problem) and the magnitude of the variations of the flow field in order to produce smooth vector  fields. One of the limitations of the method is that, typically, it can only estimate small motions. In the presence of large displacements, this method fails when the gradient of the image is not smooth enough.

\begin{figure}
\centering
\includegraphics[width=6.cm]{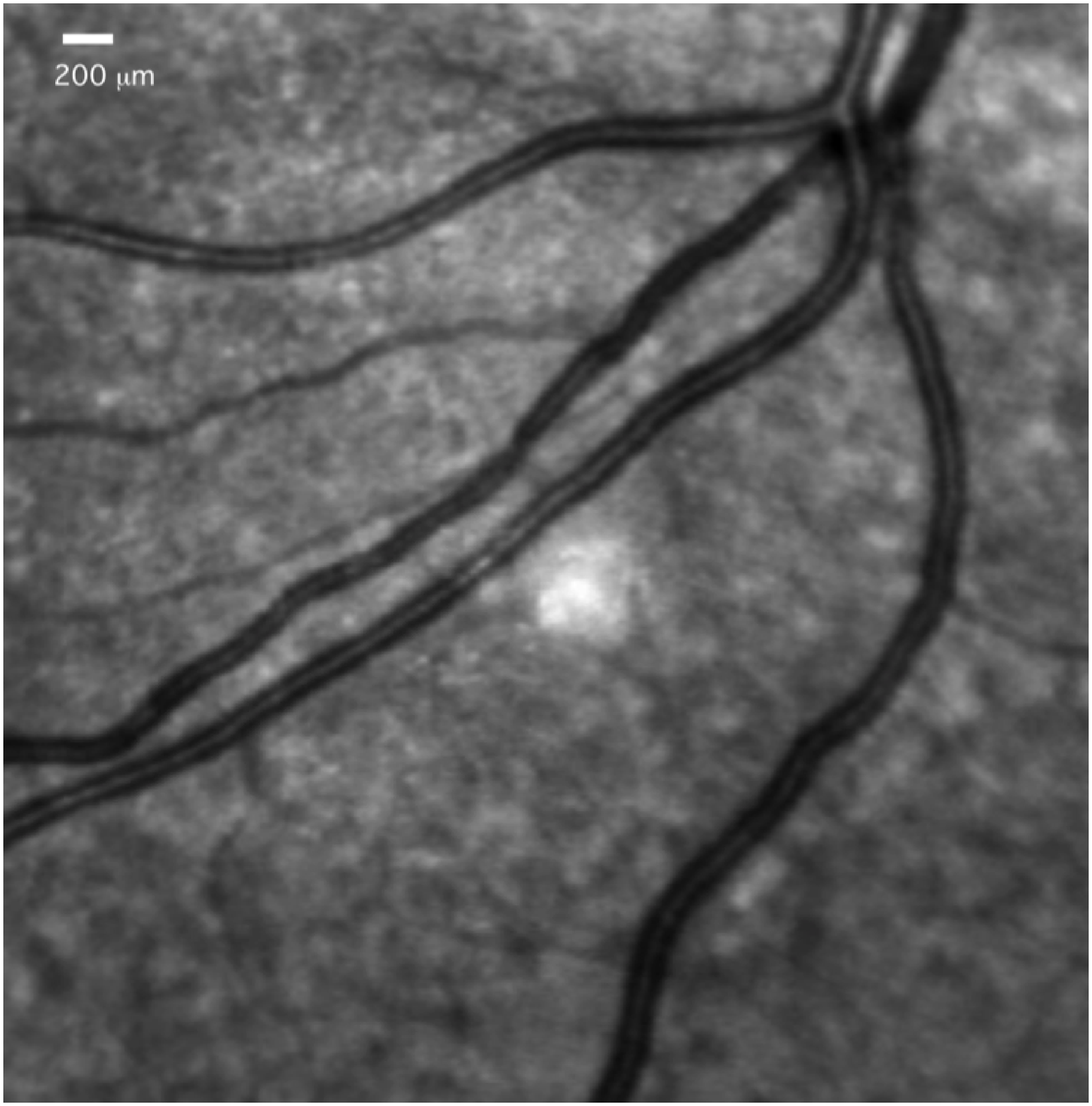}
\includegraphics[width=6cm]{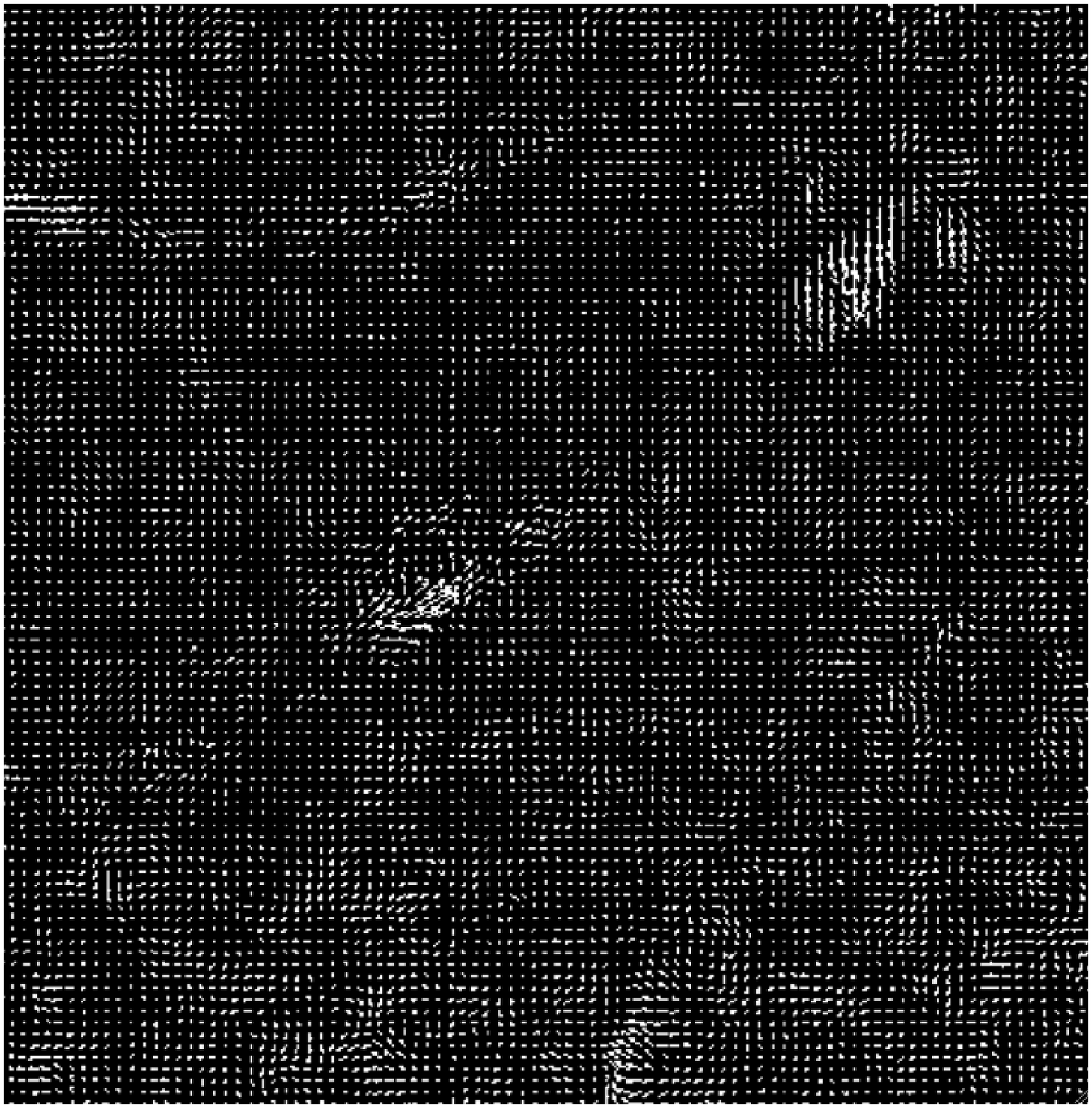}
\caption{Determination of the optical flows on a SD-OCT infrared fundus image sequence at the Ichilov hospital, Tel Aviv (left on the figure) with a current imaging Heidelberg system. The white mark indicates  a 200 $\mu$m size on the retina. The corresponding vector map is displayed right. Heart beats are determined with a sensitivity of 40 $\mu$m (smallest arrows) during the temporal sequence (160 images at a frame rate of 9 images per second) shown on the movies.   }
\label{csoct}
\end{figure}

We successfully applied the NLR method to determine the optical flow and the moving direction vectors using a current SD-OCT infrared imaging system at the Ichilov Hospital in Tel Aviv (Heidelberg infrared reflectance imaging flash fundus observed at 870 nm with a resolution of 11 $\mu$m and a frame rate acquisition of 9 images per second). Fig \ref{csoct} presents one image  extracted from the entire sequence shown in the movie.  The NLR method is applied on the entire sequence made up 160 images. Then, we applied a $k - \omega$ filter to cut-off low frequencies in the Fourier domain to compensate for the intensity fluctuations \cite{Title89,Molodij10}. We follow proper motions, distinguishing the artifacts due to the intensity variation from the optical flow. In this set of data, we are not able to follow the blood cell motions inside the arteries or the veins. We show that the subjective impression of motions inside the veins or arteries come from the variations of luminosity during the measurements. The proper motions derived from the structures are compatible with heart beats (10 beats along the sequence) in agreement with the frame rate cadence of the detector.  The optical flow accuracy is validated by the sub-pixel registration reliability (about 11 $\mu$m on the field).

\section{Discussion}

The principal purpose of the NLR method is to obtain an image quality on a sufficiently important field of view to envision medical diagnostics using current instrumentation available at the hospitals. Nevertheless, we find that the efficiency of the method can only be properly evaluated with high spatial resolution images. 

Another interest of the proposed method is to reduce the processing time (around ten times by comparison with the linear registration). The entire processing time cost is less than one second per image with a current MacBook Pro. The slow processing of the linear register method is essentially due to the rigid rotation compensation on each image of the sequence. Moreover, the linear registration routine has to detect the center of the rotation of the region on interest and evaluate the registration for several small values of the image rotation while the NLR method permits the rotation compensation on the entire field of view from LCT measurements. The registration of high spatial resolution images is accompanied inevitably with large eye saccades. The template changes very frequently despite the adaptive optics compensation. Selecting the sharpest image of the sequence as the reference image can easily done using an absolute assessment such as the entropy. Nevertheless, the optimization of the registration method is obtained when determining the best template on the basis of the frequency of occurrence that would be based on the image recognition technique and not only on the sharpest image criterion. The velocity map to describe the optical flow must then take into account the selective step for the determination of the temporal gaps. The SIA criterion becomes of high interest for selection and registration of similar images from the long sequence.

The nonlinear image registration is equivalent to optical flow when applied to a consecutive frames for image registration. The NLR method distinguishes the eyes saccades from the feature proper motions. A relatively long temporal sequence is then a necessary condition to weight the contribution of the distortion in the image in order to remove large eye saccade displacements from the smooth drifts in respect with the sampling. Temporally averaging the time series of displacement maps does not add the contribution from the saccade noise in an optimum way when saccades are variable. 

Last difficulty remains on the choice of the image quality assessment to evaluate the image quality. In the paper, we investigate the use of metrics which were originally proposed for astronomical adaptive optics imaging. We find that entropy assessment seems to match better with our subjective estimation of the quality, showing interesting close linear property in respect with the image corrugation. The entropy assessment is easy to apply and is a robust criterion when significant changes of the template occur.

\section{Conclusion}

A new data processing technique has been developed to register a sequence of extended images of  the retina. The proposed NLR procedure is able to superpose the images with a precision less than the resolution of the detector element and to resolve structures of the size of the cells of the retina on large field of views. We describe a technique that permits precise measurements of the proper motion of tracers seen on successive images of a time serie. The cross correlation is defined as a  function of position in the image, within a spatially localized apodized window. The time average of the spatially localized cross correlation gives measure of the displacement that is not biased by eyes saccades.  Many biological tasks can use future image processing techniques for fully automatic image analysis. This study is applying the knowledge acquired in the fields of optics and imaging in astrophysics in order to improve the retinal imaging at very high spatial resolution in order to perform a medical diagnosis for diabetic retinopathy, HIV/AIDS-related retinitis and age-related macular degeneration, conditions that can lead to blindness without early diagnosis and treatment. Diagnosing diseases that affect the retina requires highly trained specialists who use equipment to visualize the inner parts of the eye.

\section*{Acknowledgements}
We would like to thank Pr. A. Loewenstein, director of the Ophtalmology division and Pr. A. Barak of the Tel Aviv Sourasky medical center for supplying us with the wide retinal images and  for providing very constructive comments.

\end{document}